\documentclass[aps,pre,twocolumn, fleqn,superscriptaddress,showpacs,floatingfix,nofootinbib]{revtex4-1}
\usepackage{epsf,amsmath,amssymb,verbatim,color,multirow,pifont,hyperref,soul}
\usepackage{xcolor}
\usepackage{graphicx}
\usepackage{grffile}
\usepackage{subfigure}
\newcommand{\spe}{(s)}
\newcommand{\rea}{(r)}

\begin{document}

\title{Essentiality landscape of metabolic networks}
\author{P. Kim}
\affiliation{CCSS and CTP, Department of Physics and Astronomy, Seoul National University, Seoul 08826, Korea}
\author{K. Han}
\affiliation{Department of Economics, University of Michigan, Ann Arbor, Michigan 48109, USA}
\author{D.-S. Lee}
\email{deoksun.lee@inha.ac.kr}
\affiliation{Department of Physics, Inha University, Incheon 22212, Korea}
\author{B. Kahng}
\email{bkahng@snu.ac.kr}
\affiliation{CCSS and CTP, Department of Physics and Astronomy, Seoul National University, Seoul 08826, Korea}

\begin{abstract}
Local perturbations of individual metabolic reactions may result in different levels of lethality, depending on their roles in metabolism and the size of subsequent cascades induced by their failure. 
Moreover, essentiality of individual metabolic reactions may show large variations within and across species. Here we quantify their essentialities in hundreds of species by computing the growth rate after removal of individual and pairs of reactions by flux balance analysis.  We find that about $10\%$ of reactions are essential, i.e., growth stops without them, and most of the remaining reactions are redundant in the metabolic network of each species. This large-scale and cross-species study allows us to determine {\it ad hoc}  ages of each reaction and species. We find that when a reaction is older and contained in younger species, the reaction is more likely to be essential. Such correlations of essentiality with the ages of reactions and species may be attributable to the evolution of cellular metabolism, in which alternative pathways are recruited to ensure the stability of important reactions to various degrees across species.
\end{abstract}

\maketitle

\section{Introduction}

%%%% Structure determines function
The persistent generation and consumption of matter and energy by numerous cellular components despite internal and external perturbations are crucial to the survival and reproduction of all living organisms. The topology and dynamics of cellular networks afford this stability as demonstrated in previous works about the effects of network structure on the functionality and regulation of cellular networks~\cite{Stelling2002Nature,Jain13PRE,Shlomi2005PNAS}, the functional plasticity and redundancy of metabolic networks for stability after environmental changes~\cite{harrison2007PNAS,Almaas2005Plos,Wagner2005BioEssays,Serrano2014plosone}, the effect of environmental condition on gene and reaction essentiality~\cite{Suthers2009Mol,Barve2012PNAS,Joyce2006JBac,Pal2006Nature} and maintenance of stability after multiple knockouts~\cite{deutscher2006genetics,papp2004nature,Nakahi2009Mol,Matias2009Plos}.

Metabolic networks represent collections of a large number of reactions converting numerous metabolites toward generation of biomass and energy. Because a metabolite can be processed by multiple reactions and a reaction involves multiple metabolites, the topology of metabolic networks is complicated. Although they are generally stable, some reactions and metabolites may yield abnormal fluxes and concentrations owing to stochasticity and perturbations. Local perturbations may lead to system-level malfunctioning via cascades of failures. In contrast, inactivation of some reactions may not be lethal if alternative pathways can replace the functions that the inactivated reactions had been performing. Given that individual metabolic reactions play a variety of roles specified by their physical, chemical, and biological characteristics and thus have different importance and essentiality, understanding which reactions may cause global damage in a given metabolic network is an essential step toward understanding the stability of cellular networks.    

Studying the network level stability of cellular metabolism in a single species such as {\it Escherichia coli}~\cite{Stelling2002Nature,Joyce2006JBac,palsson13Science} or yeast~\cite{harrison2007PNAS,Serrano2014plosone,deutscher2006genetics,papp2004nature} may be insufficient. Cellular metabolism has evolved across species on a long term scale, developing various pathways and thereby enhancing stability~\cite{Light2004sp,Yamada2009qe,KimSciRrep2015}. A metabolic network of a single species may be only one of many possible implementations of metabolism that evolves differently contingent upon the perturbations imposed on the species. Thus, failure of the same reaction or at the same metabolite in a network may yield different lethality in  different species. In this regard, we computed the effects of inactivation of each single reaction and each pair of reactions within the framework of flux balance analysis (FBA)~\cite{palsson2006systems,segre2002PNAS,palsson2014Nature} to quantify their essentiality rates in hundreds of species. The metabolic networks of those species are obtained from the BioCyc database~\cite{Karp01012005}. Each reaction is classified as {\it essential, active, or redundant} depending on its contribution to biomass generation. Moreover, a pair of reactions are regarded as a {\it backup pair} for each other if their synthetic contribution is larger than the sum of the individual ones. 

The distribution of such essentiality grades over reactions and species is far from random, revealing the evolutionary pressures imposed on the metabolism. Defining the popularity of a reaction as its evolutionary age and the proportion of young, less popular, reactions in a species as its species age, we find that a reaction is more likely to be essential when its age is older and when it is present in a younger species. Given that the essentiality of a reaction depends on its intrinsic physicochemical features and on its connection to other reactions and metabolites; this result is an evidence that cellular metabolism has evolved toward enhancing stability by recruiting alternative metabolic pathways against diverse perturbations. Our study reveals that large-scale cross-species research can greatly contribute to understanding the evolution of cellular metabolism from the standpoint of their stability subject to a changing environment.

%\section{Results}

\section{Measure of Essentiality grade and backup reactions}
\label{sec:lethaldef}
We consider a total of $N_R=6911$ distinct reactions that are found in at least one of $N_S=386$ species.  These species were selected because they contain a sufficient number of reactions and biomass components satisfying the criterion introduced in Appendix A. Let us denote $N_R(s)$ as the number of reactions contained in species $s$. For each reaction $r_i(s)$ ($i=1,\cdots, N_R(s)$) of species $s$ ($s=1,\cdots, N_S$), we determine its essentiality grade $G_{r_i}(s)$ as one of five discrete levels i)$-$ v) defined below.  To perform this task, we selected 129 core biomass components indexed as $b_i=1,\cdots, N_{B}=129$ from six species whose metabolic networks are extensively investigated as stated in Appendix B. Those core biomass components serve as the bases for determining the essentiality grade of each reaction.  

For each species $s=1,\cdots, N_S$, using FBA, we check whether each biomass component $\{b_i\}$ is produced or not. Then we let only the produced components participate in the species-specific biomass synthesis reaction. For instance, if  species $s$ produces biomass components $b_3, b_7$, and $b_{41}$, then the biomass synthesis reaction is given as $b_3 + b_7 + b_{41} \xrightarrow{g_{w}(s)} biomass$, where $g_{w}(s)$ is the flux of the synthesis reaction and equivalent to the growth rate of that species as a wild type. To determine essentiality grade $G_r(s)$ of individual reaction $r$ for certain species $s$, we remove reaction $r$ and then calculate the growth rate by FBA, and the result is denoted as $g_{r}(s)$.
\begin{itemize}
	\item[i)] If $g_{r}\spe=0$, reaction $r$ is regarded as {\it essential} and its essentiality grade is given as $G_{r}(s)=\mathcal{E}$: Inhibition of reaction $r$ prevents species $s$ from growing.  
	\item[ii)] If $g_{r}(s)\ne 0$ but $g_{r}(s) < g_{w}(s)$, reaction $r$ is regarded as {\it active} and the essentiality grade is expressed as $G_{r}(s)=\mathcal{A}$: Inhibition of reaction $r$ reduces the growth rate but is not so lethal that the organism would stop growing. 
	\item[iii)] If $g_{r}\spe=g_{w}\spe$, the reaction $r$ is regarded as {\it redundant} ($G_{r}\spe=\mathcal{R}$): Biomass generation is not affected by the inhibition of reaction $r$. 
	\item [iv)] If reaction $r$ is present but not connected to any biomass component, it is regarded as {\it isolated} ($G_{r}\spe=\mathcal{I}$). 
	\item[v)] If reaction $r$ is absent in species $s$, its essentality grade is set to {\it nonpresent} ($G_{r}(s)=\mathcal{N}$), i.e., absent. 
\end{itemize}

%%%%% Fig. 1 fig:Comparegrade %%%%%%%%%%%%
\begin{figure}
\centering
\includegraphics[width=0.95\linewidth]{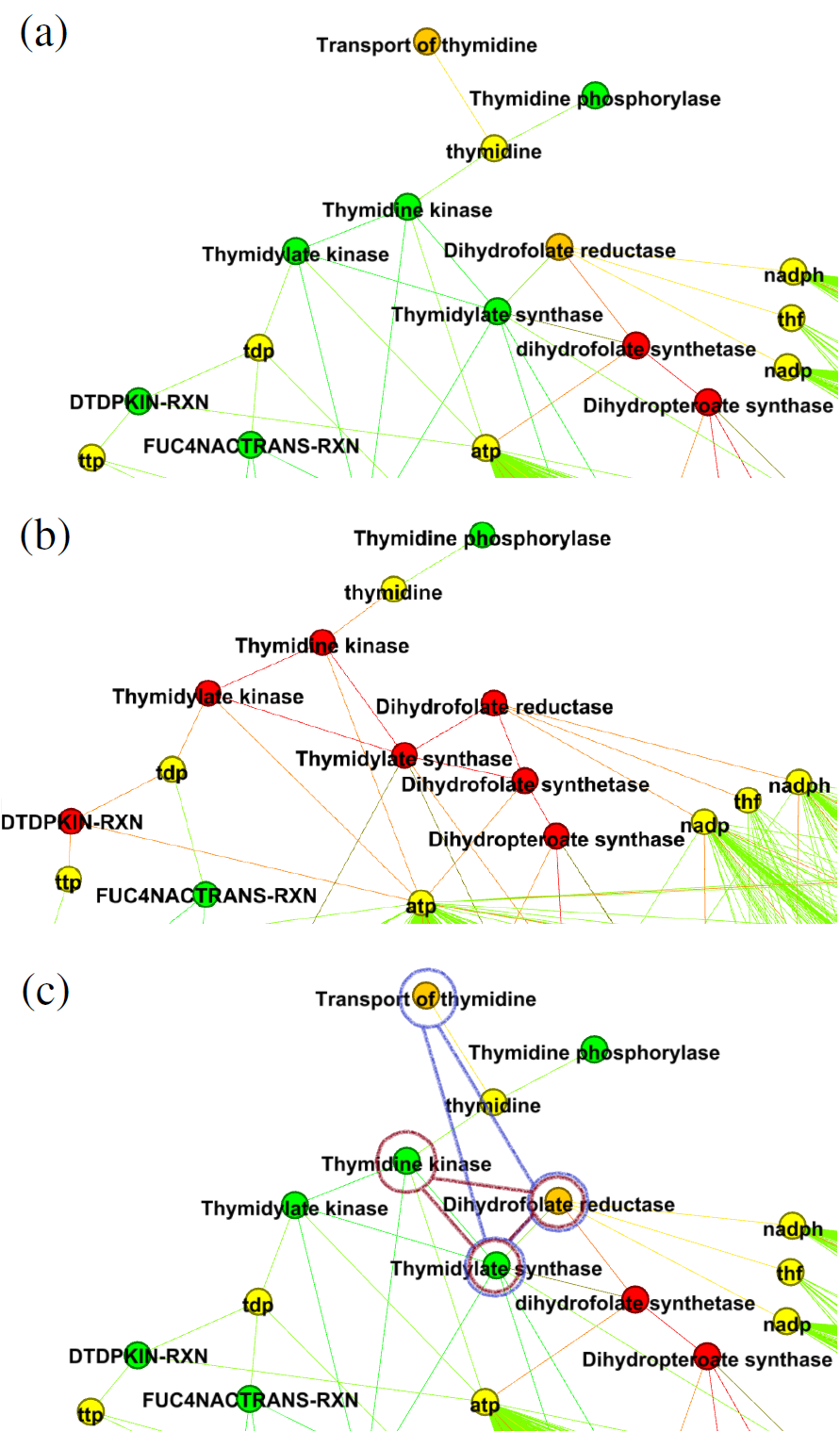}
\caption{Examples of essentiality grades and backup relation in metabolic networks. Essentiality grades of some reactions in (a) {\it E. coli} and (b) {\it A. tumefaciens C58}. Nodes represent reactions and are connected when they share at least one metabolite that are shared by no more than four distinct reactions. Essential $(\mathcal{E})$, active  $(\mathcal{A})$, and redundant $(\mathcal{R})$ reactions are denoted as red, orange and green balls, respectively. Biomass components are yellow. 
(c) Backup relation in the metabolic network of {\it E. coli}. Each pair of nodes surrounded by circles of the same color (purple or violet) is subject to a backup relation. For instance, any two of thymidine kinase, thymidylate synthase, and dihydrofolate reductase are a backup pair. The same holds for transport of thymidine, thymidylate synthase, and dihydrofolate reductase.}
%\label{fig:ecoliBackup}
\label{fig:Comparegrade}
\end{figure}
%%%%%%%%%%%%%%%%%%%%%%%%%%%%%%%%%

Figs.~\ref{fig:Comparegrade}(a) and \ref{fig:Comparegrade}(b) show the essentiality grades of some reactions in two species. Dihydrofolate reductase, which reduces dihydrofolic acid to tetrahydrofolic acid, is active in {\it Escherichia coli} but essential in {\it Agrobacterium tumefaciens C58}. Different grades are also seen for thymidine kinase, thymidylate synthase, and thymidylate kinase, which are redundant in {\it E. coli} but essential in {\it A. tumefaciens C58}. 

The reduction in growth rate $g_{w}(s)-g_{r}(s)$ caused by inhibition of a certain reaction $r$ can be a measure of its global damage. Similarly, we consider the reduction in growth rate $g_{w}\spe-g_{r,r'}\spe$ caused by the simultaneous inhibition of two reactions $r$ and $r'$  to determine whether they cooperate as a backup pair. If reactions $r$ and $r'$ contribute independently to biomass generation,  then $g_{w}\spe-g_{r,r'}\spe$ would be equal to the sum of individual damages $g_{w}\spe-g_{r}\spe$ and $g_{w}\spe-g_{r'}\spe$. Nontheless, if $g_{w}\spe-g_{r,r'}\spe$ is greater than $(g_{w}\spe-g_{r}\spe) + (g_{w}\spe-g_{r'}\spe)$, the inhibition of either $r$ or $r'$ alone is not as lethal due to the enhanced contribution of the remaining reaction $r$ or $r'$, respectively.  This dependence can reduce the damage compared to that without the enhancement; the two reactions are not independent but back each other up. Therefore, we regard two reactions $r$ and $r'$ as backup if they satisfy 
\begin{equation}
g_{w}\spe+g_{r,r'}\spe - g_{r}\spe-g_{r'}\spe<0.
\label{eq:bp}
\end{equation}
Eq.~(\ref{eq:bp}) indicates that two backup reactions can partially substitute for each other in their overall contribution to biomass generation. Therefore, their simultaneous inactivation is more lethal than expected from individual damage. Some examples of backup reactions are given in Fig.~\ref{fig:Comparegrade}(c).

%%%%%%%%%%%%%%%%%% Fig. 2 (old fig 3) fig:gradePortion %%%%%%%%%%%%
\begin{figure*}
\centering
\includegraphics[width=0.9\linewidth]{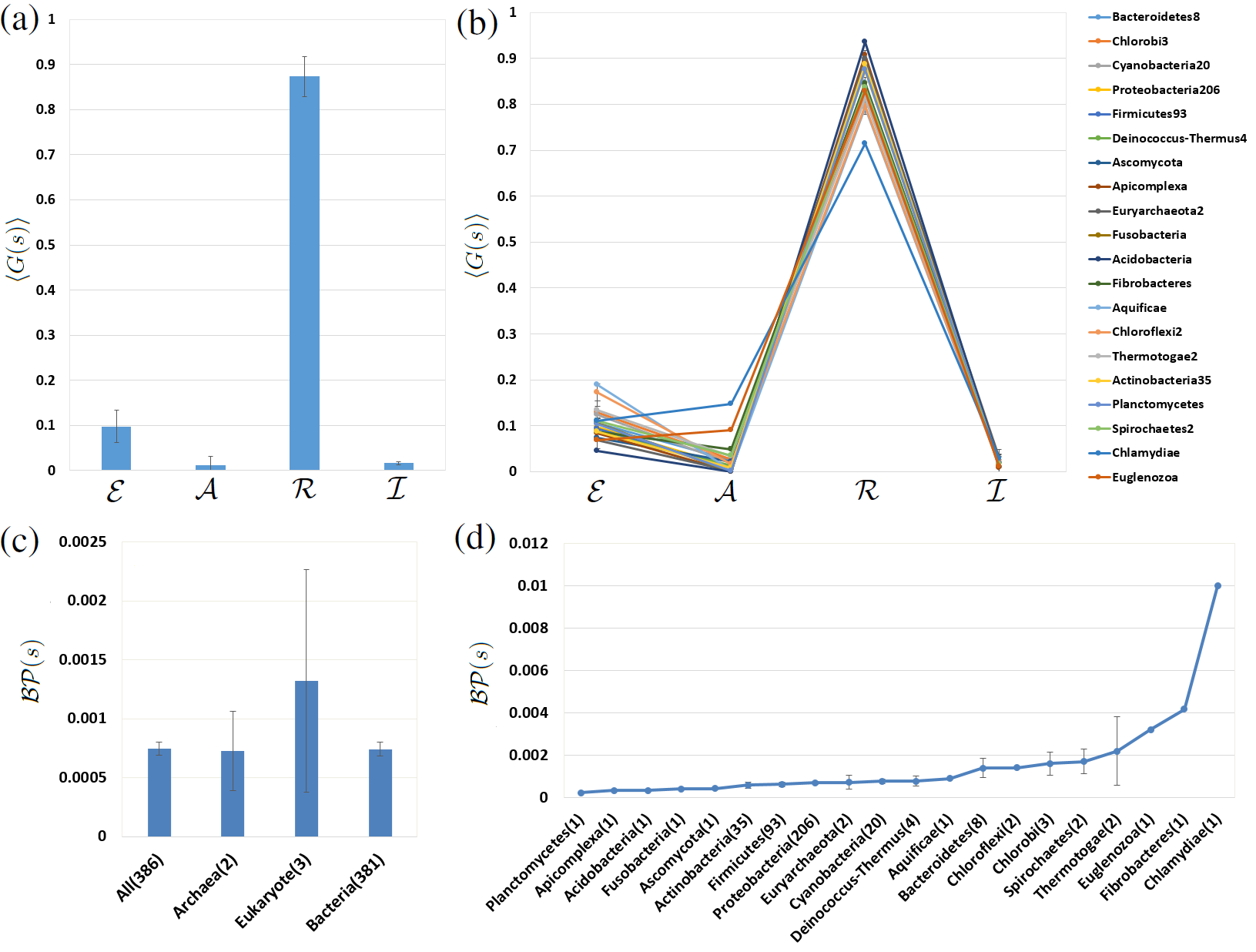}
\caption{The proportion of reactions of each essentiality grade and backup pairs in a species. 
(a) The mean and standard deviation of the proportion of each essentiality grade across species. 
(b) The proportion of each essentiality grade for species belonging to each phylum. 
(c) The proportion of backup pairs for each domain.
(d) The proportion of backup pairs for each phylum. The numbers of species belonging to each phylum are given in parentheses. Apicomplexa, Ascomycota and Euglenozoa belong to Eukarya; Euryarchaeota to  Archaea; and the rest belong to Bacteria.
}
\label{fig:gradePortion}
\end{figure*}
%%%%%%%%%%%%%%%%%%%%%%%%%%%%%%%%%%%%%%%%%%%

We find that the proportion of essential reactions per species is $\mathcal{E}(s) =9.7\pm 3.5\%$, that of active reactions is $\mathcal{A}(s)=1.3\pm 1.8\%$, the proportion of isolated reactions is $\mathcal{I}(s)=1.6\pm 0.4\%$, and that of redundant reactions is $\mathcal{R}(s)=87\pm 4.5\%$ as shown in Fig.~\ref{fig:gradePortion}(a). Their small standard deviations across species and weak variation across phyla shown in Fig.~\ref{fig:gradePortion}(b) point to the robustness of these proportions. A large proportion of redundant reactions may be related to the reported redundancy of many metabolism-related genes in yeast~\cite{papp2004nature,Wagner2005BioEssays,Wang2009Genome}.

The proportion of backup reaction pairs is $\mathcal{B_P}(s)=0.074 \pm 0.0056\%$.
Most species studied in this paper belong to the bacteria domain and  only two species to Archaea and three to Eukarya, but the proportion of backup pairs significantly varies from domain to domain: Archaea have the smallest proportion of backup pairs, and Eukarya have the largest proportion [Fig.~\ref{fig:gradePortion}(c)]. In this regard, it has been argued that Archaea live in extreme and narrowly defined environments and thus do not need to have large sets of backup pairs~\cite{Borenstein2008PNAS,Ebenhoh2005Genome}. 
The variation of the proportion of backup pairs across phyla is presented in Fig.~\ref{fig:gradePortion}(d).

%%%%%%%%%%%%%%%%%% Fig. 5 fig:gradedistribution %%%%%%%%%%%%

%%%%%%%%%%%%%%%%%%%%%%%%%%%%%%%%%%%%%%%%%%

\section{Evolutionary ages of reactions and species}

%%%%%%%%%%%%%%%%%% Fig. 3 fig:P(N_R(s)) %%%%%%%%%%%%
\begin{figure}
\centering
\includegraphics[width=0.95\columnwidth]{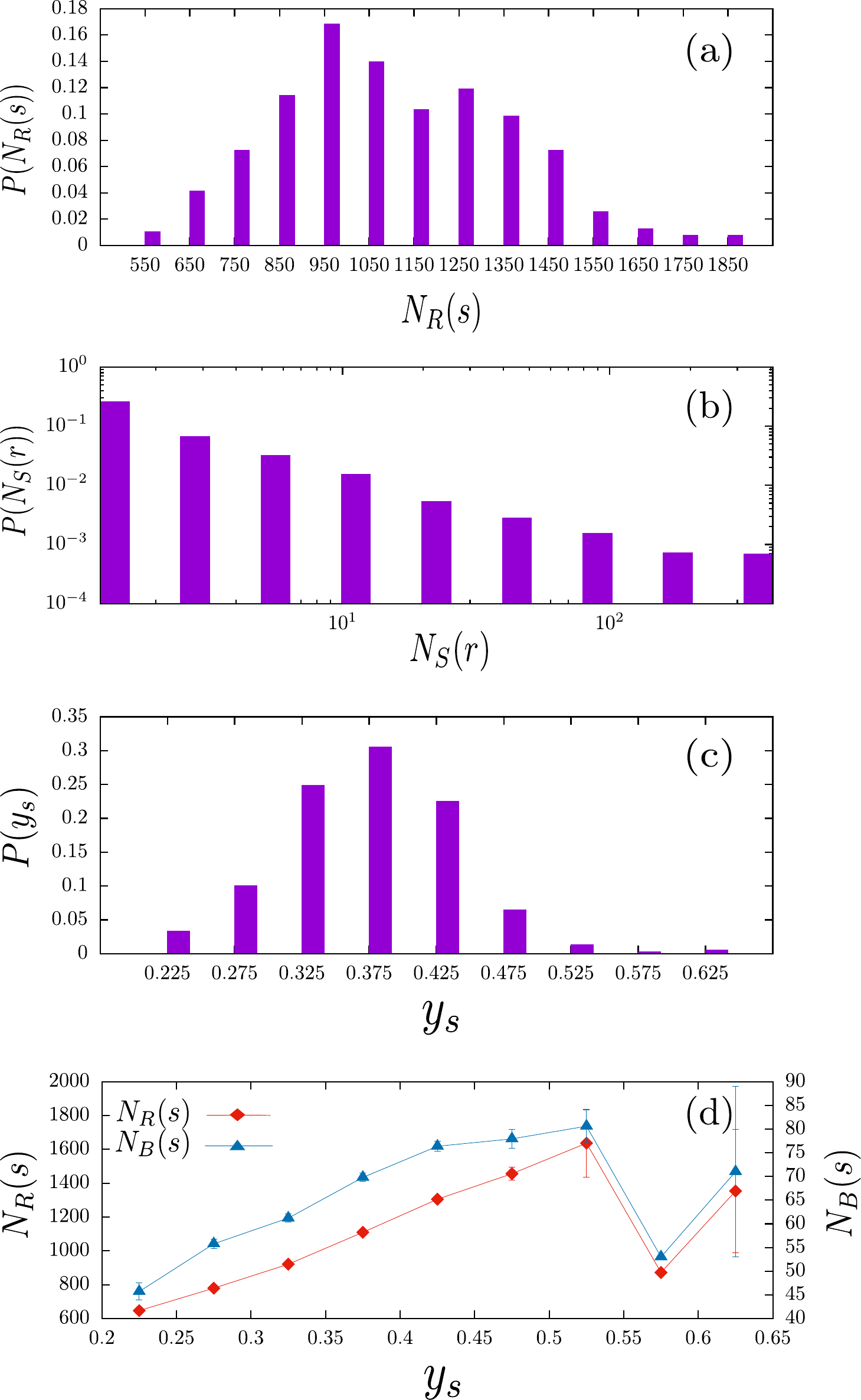}
\caption{Evolutionary ages of reactions and species. (a) The proportion of the species containing $N_R(s)$ reactions: The mean and standard deviation are $1090$ and $260$, respectively. 
(b) Distribution of the number of species $N_S(r)$ containing each reaction $r$ in log scales. Note that $N_S(r)$ is related to the age of reaction $r$ as $N_S(r) = y_r N_S$, where $N_S$ is the total number of species considered in this work. (c) Distribution of species ages $y_s$. Nine bins of size $0.05$ divide the whole range of $y_s$ between $0.200$ and $0.650$. Quite low densities at the two rightmost bins indicate that only a few species have $y_s\geq 0.5$, which induces significantly large error bars for later data analyses. (d) The number of reactions per species $N_R(s)$ (left) and  the number of biomass components per species $N_B(s)$ (right) are given as a function of species age.
}
%\label{fig:fraction_n_r}
\label{fig:age}
\end{figure}
%%%%%%%%%%%%%%%%%%%%%%%%%%%%%%%%%%%%%%%%%%%

To introduce the ages of reactions and species, we first consider distribution $P_S(N_R(s))$ of the number of reactions $N_R(s)$ over species $s$. The distribution shown in Fig.~\ref{fig:age}(a) is centered on the mean value about $1090$, implying that the numbers of reactions present in different species are almost homogeneous around the mean value. Next, we consider the number of species $N_S(r)$ that contain reaction $r$ in their metabolism. After that, we count the number of reactions $N_R(N_S(r))$ that belong to $N_S(r)$ distinct species; this relation is shown as a function of $N_S(r)$ in Fig.~\ref{fig:age}(b). It follows a broad distribution, which implies that a few reactions belong to a large number of species, whereas many other reactions are present in only a few species. Such heterogeneous frequencies may be rooted in the difference of time points when distinct reactions were introduced into the metabolic networks of certain living species. The reactions that appear in many species are expected to play a central role in the metabolism since the early period of its evolution. In contrast, reactions that appear in a few species may have been introduced only recently to perform functions required by selected species and their environments. Therefore, the number of species $N_S\rea$ that contain a certain reaction $r$ in their metabolism divided by the total number of species studied in this work $N_S$ may represent the evolutionary age of reaction $r$, denoted as $y_r$, i.e., 
\begin{equation}
y_r=\frac{N_S\rea}{N_S}.
\label{eq:yr}
\end{equation}
Thus, Eq.~(\ref{eq:yr}) implies that old reactions were inherited by many contemporary species. 

If reactions are introduced to cope with environmental requirements, the distribution of reaction age in a species may depend on the environments of its ancestors and the metabolic evolution realized in the lineage. If a species has mostly old reactions and only a few young reactions, then the species may have had insufficient time to enrich its metabolism by recruiting new reactions or little motivation to do so owing to a rich environment. Therefore, how young the reactions present in a species are may represent the degree of metabolic evolution or evolutionary age of the species. We define evolutionary age $y_s$ of species $s$ as 
\begin{equation}
y_s=\frac{\sum_{r \in \mathbb{R}(s) } (1-y_r)}{N_R(s)},
\label{eq:ys}
\end{equation}
where $\mathbb{R}(s)$ is the set of reactions contained in the metabolism of species $s$. $1-y_r$ is the proportion of species that do not have reaction $r$; and $N_R(s)$ is the number of reactions present in  species $s$. Consequently, old species with large $y_s$ contain many young reactions;
they have evolved considerably to possess many young, and possibly noncore reactions that play specific roles required by their surroundings. The distribution of species age is given in Fig.~\ref{fig:age}(c). The populations of species at a middle age are dominant. We find that {\it Trypanosoma brucei} and {\it E. coli} are the oldest among the species we studied, whose $y_s$ are 0.63 and 0.62, respectively, whereas {\it Wolbachia pipientis} and {\it Ehrlichia ruminantium Welgevonden} are the youngest with ages $0.21$ and $0.22$, respectively. 

%%%%%%%%%%% new fig. 6 %%%%%%%%%
%\begin{figure}
%\centering
%\includegraphics[width=0.4\columnwidth]{fig6_New.png}
%\includegraphics[width=0.7\columnwidth]{fig6_New_line.png}
%\caption{The number of reactions and biomass components depending on species age. (a) The number of reactions per species as a function of species age. (b) The number of biomass components per species as a function of species age. 
%}
%\label{fig:nrsnbms}
%\end{figure}
%%%%%%%%%%%%%%%%%%%%%%%%%

We remark that the species age defined here and even the order among different species may not be consistent with the real age of species in the phylogenetic tree of life but represents the degree of evolution of the metabolic reactions. Nevertheless as shown later, various properties of cellular metabolism manifest strong dependence on their species age in a systematic manner, suggesting that this {\it ad hoc} species evolutionary age may be a reasonable representation for evolutionary studies.

In Fig.~\ref{fig:age}(d),  the number of reactions $N_R(s)$ contained in species $s$ tends to increase with the species age $y_s$ as expected from our definitions of reaction age and species age. Whereas, old reactions appear in all species regardless of the number of reactions present in them, young reactions appear only in the species having a large number of reactions, which are assigned large values of $y_s$ according to Eq.~(\ref{eq:ys}). The Pearson correlation coefficient (PCC) between the average number of reactions $\langle N_R(s)\rangle_{y_s}$ for the species of age $y_s$ and age $y_s$ is as large as $0.86$. Similarly, the number of biomass components $N_{B}(s)$ available in each species increases with the species age with the PCC $=0.68$ as shown in  Fig.~\ref{fig:age}(d). This correlation may originate in the strong correlation between $N_{B}(s)$ and $N_R(s)$ (PCC $=0.84$).

%%%%%%%%%%%%%%%%%% Fig. 4 fig:gradedistribution %%%%%%%%%%%%
\begin{figure}
\centering
\includegraphics[width=0.9\columnwidth]{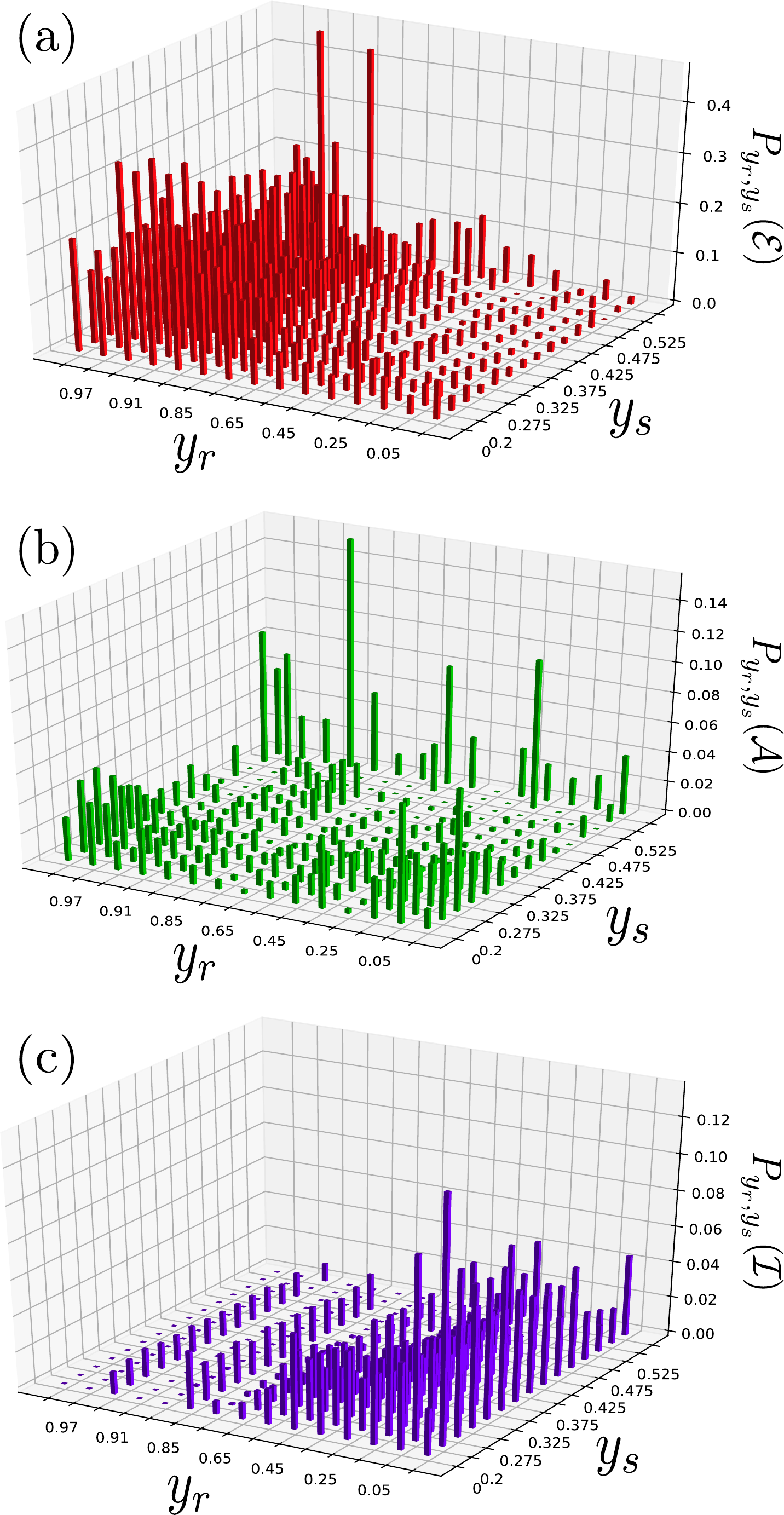}
\caption{
(a) Probability $P_{y_r,y_s}(\mathcal{E})$ of a reaction of age $y_r$ in a species of age $y_s$ to be essential.  
(b) Probability $P_{y_r,y_s}(\mathcal{A})$ to be active. (c) Probability  $P_{y_r,y_s}(\mathcal{I})$ to be isolated. 
}
\label{fig:gradeDistribution}
\end{figure}
%%%%%%%%%%%%%%%%%%%%%%%%%%%%%%%%%%%%%%%%%%%

\section{The Essentiality landscape on the species- and reaction-age plane}

How the essentiality grade of a reaction depends on its reaction age and the age of species that it is contained in can be understood by drawing essentiality landscapes [Fig.~\ref{fig:gradeDistribution}], in which probabilities $P_{y_r, y_s}(G)$ that a reaction of age $y_r$ present in a species of age $y_s$ has grade $G=\mathcal{E}, \mathcal{A}$ and $\mathcal{I}$ are shown. It is noteworthy that probability $P_{y_r,y_s}(\mathcal{E})$ to be essential appears large in the region of old reaction age, particularly, in the region of young ages of species as shown in Fig.~\ref{fig:gradeDistribution}(a). Probability $P_{y_r, y_s}(\mathcal{A})$ of the active grade is distributed over the entire region in Fig.~\ref{fig:gradeDistribution}(b). Probability $P_{y_r,y_s}(\mathcal{I})$ for the isolated grade becomes large in the region of young reaction age and species age as depicted in Fig.~\ref{fig:gradeDistribution}(c), respectively. 
Such correlations between the essentiality grades and ages of reaction and species can be further analyzed by projecting them onto the space of species age or reaction age only, as discussed below, providing deep insights into the evolutionary origin of the essentiality landscape illustrated in Fig.~\ref{fig:gradeDistribution}.

\subsection{Popularity of essential reactions in young species}

%%%%%% new fig 5 (old Fig 7) Stablity of a species depending on the age %%%%%%%%%%%%%%%
\begin{figure}
\centering
\includegraphics[width=1.0\columnwidth]{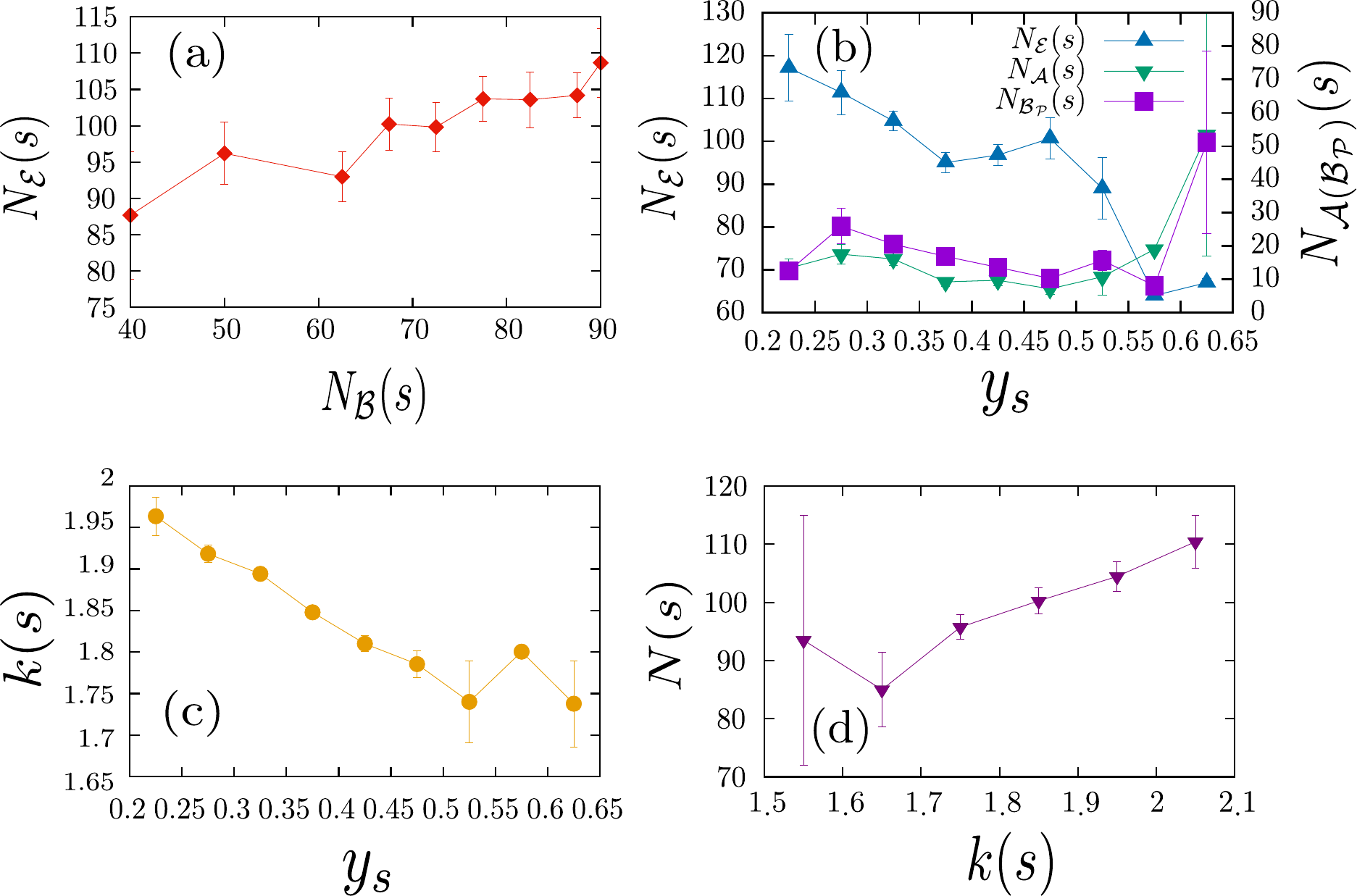}
\caption{Essentiality and backup availability of a metabolic network versus species age.  
(a) Plot of the number of essential reactions per species versus the number of biomass components. They correlate positively (PCC = $0.188$, P$=0.0002$). (b) The numbers of essential reactions $N_\mathcal{E}(s)$, active reactions $N_\mathcal{A}(s)$, and backup pairs $N_{\mathcal{B_P}}(s)$ per species versus $y_s$. The $y$ axis (left) is for $N_\mathcal{E}(s)$ and that (right) is for $N_\mathcal{A}(s)$ and $N_{\mathcal{B_P}}(s)$. The $y$-axis scale (right) for $N_{\mathcal{B_P}}(s)$ should be multiplied by $20$. Data of $N_\mathcal{E}(s)$ and $N_{\mathcal{B_P}}(s)$ show significant negative correlations with $y_s$, characterized by PCC$=-0.22$ and $-0.11$ (P=0.032), respectively. The correlation between $N_\mathcal{A}(s)$ and $y_s$ is not significant; PCC= $-0.09$ (P=0.06). 
(c) $k(s)$ denotes global connectivity averaged over the species within interval $[y_s-\delta, y_s+\delta]$, where $2\delta=0.05$ is taken. Plot of the averaged global connectivity of the metabolic reaction network as a function of species age $y_s$. Their correlation  is given by PCC = $-0.506$. 
(d) Plot of the number of essential reactions per species  versus global connectivity $k(s)$.  Their PCC is  $0.204$.
}
\label{fig:bm_ess}
\end{figure}
%%%%%%%%%%%%%%%%%%%%%%%%%%%%%%%%%%%%

The number of essential reactions in the metabolism of a species can be an indicator of the vulnerability of the metabolism to perturbations. Along this line of reasoning, it would be interesting to identify the determinants of the number of essential reactions. Given that the essentiality of a reaction is determined by its contribution to the biomass production, a species with more  biomass components is likely to have a larger number of essential reactions (PCC $=0.188, {\rm P}=0.0002)$ as shown in Fig.~\ref{fig:bm_ess}(a). Moreover, older species have more reactions and more biomass components [Fig.~\ref{fig:age}(d)]. Thus, one could expect the older species to have  a larger number of essential reactions, if only the number of biomass components were responsible for the variation in the number of essential reactions across species. Counterintuitively, this is not the case; it is illustrated in Fig.~\ref{fig:bm_ess}(b) that older species tend to have smaller numbers of essential reactions (PCC=$-0.22$). 

The major difference between old and young species may be found in the distribution of reaction ages in their metabolic networks. A certain reaction is essential if it is involved in biomass generation pathways and its absence is not backed up by alternative pathways. We conjecture that young reactions, which are abundant in old species according to Eq.~\eqref{eq:ys}, tend to supplement old reactions, preventing inactivation of the latter from being lethal: Young species, which have a small number of young reactions, may have had insufficient time to develop such alternative pathways and are thus less stable than old species. This reasoning also suggests that metabolic networks have evolved to enhance stability by recruiting alternative pathways.

Although the number of active reactions per species does not show a significant correlation with species age as shown in Fig.~\ref{fig:bm_ess}(b), the number of backup pairs was found to be small in old species [Fig.~\ref{fig:bm_ess}(b)]. Reasoning similar to the one regarding the dependence of the number of essential reactions on the species age can be applied. If one of two reactions in a backup pair is inactivated, the other reaction plays a crucial role in diminishing the growth rate reduction. If they are simultaneously inactivated, there are no alternative pathways; therefore the reduction in the growth rate is larger than that expected from individual inactivation. Therefore, the abundance of such backup pairs in young species may be attributed to insufficient pathways.

Connecting two reactions if they process the same metabolite, when the shared metabolite is processed by no more than four distinct reactions, one can construct a network of metabolic reactions only~\cite{Lee22072008}, which represents the reaction-level pathways for perturbation spreading. Local connectivity $k_r(s)$,  the number of neighboring reactions, of reaction $r$ in the metabolic-reaction network of species $s$ and its average $k(s)$ across all reactions, which we call  the global connectivity of the species, can be therefore closely related to the stability of the metabolism of a species. Although old species have many reactions (both old and young reactions), the global connectivity was found to be smaller in older species than in younger species as presented in Fig.~\ref{fig:bm_ess}(c). The number of essential reactions per species positively correlates with the global connectivity as shown in Fig.~\ref{fig:bm_ess}(d).

%%%%%%%%%%%%%%%%%% new fig 6 (old Fig. 9) fig: %%%%%%%%%%%%
\begin{figure}
\centering
\includegraphics[width=0.99\columnwidth]{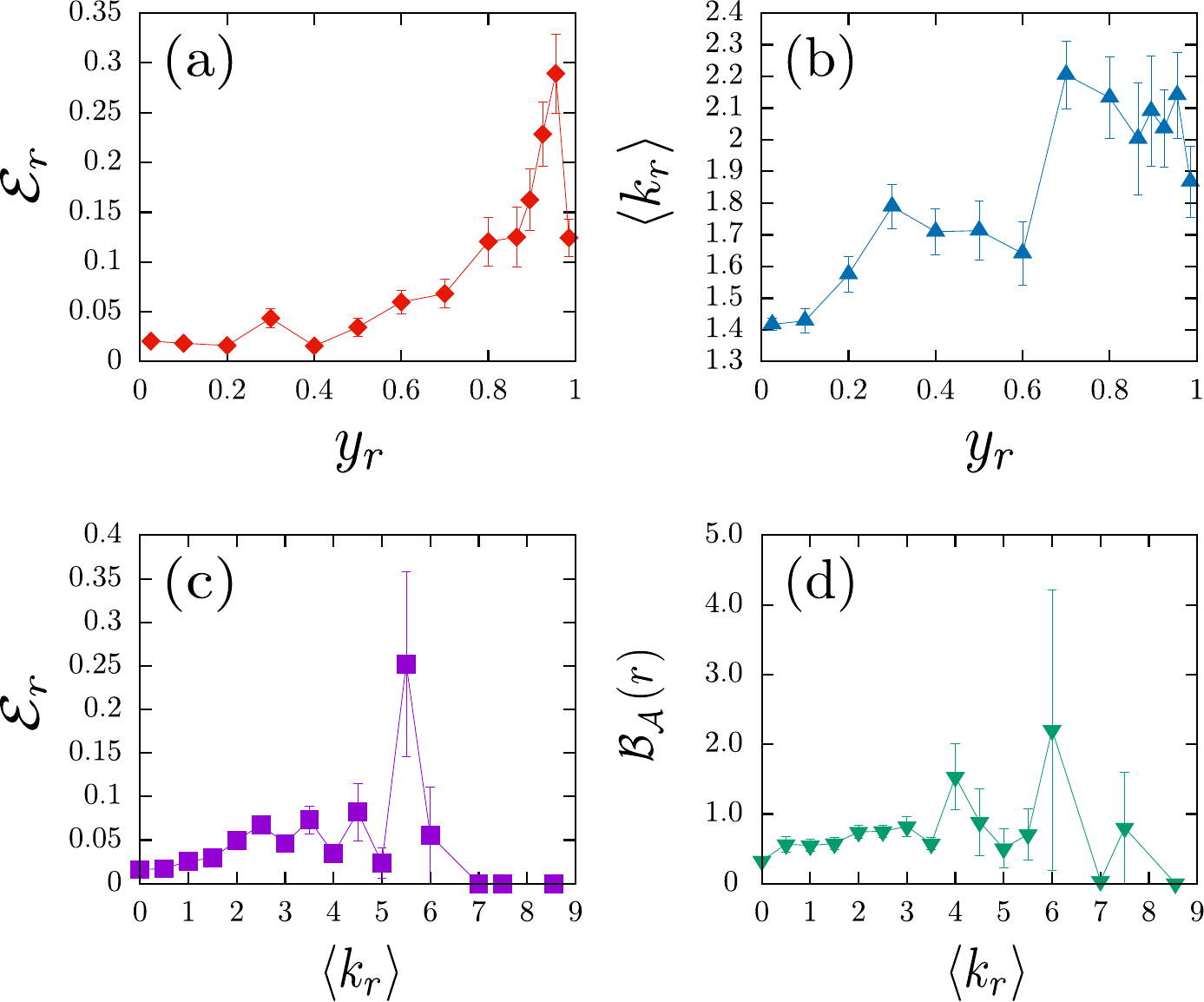}
\caption{Essentiality and backup availability across species characteristics of metabolic reactions. (a) Essentiality $\mathcal{E}_r$ of reaction $r$ as a function of its age $y_r$.  Their PCC is $0.243$. (b) Average local connectivity $\langle k_r\rangle$ versus reaction age $y_r$. Their PCC is $0.148$.
(c) Essentiality $\mathcal{E}_r$ versus average local connectivity $\langle k_r \rangle$. Their PCC is $0.103$. 
(d) Backup availability $\mathcal{B_A}(r)$ as a function of $\langle k_r \rangle$. The $y$-axis scale for $\mathcal{B_A}(r)$ should be multiplied by $10^{-3}$. Their PCC is $0.0667$. 
}
\label{fig:6}
\end{figure}
%%%%%%%%%%%%%%%%%%%%%%%%%%%%%%%%%%%%%%%%%%%

\subsection{High essentiality of old reactions in young species}

When a large number of reactions in each metabolic network is intricately wired, the essentiality of individual reactions depends on the intrinsic features of each reaction and on the organization of the metabolic network, as illustrated by our cross-species analysis. 
We introduce likelihood $\mathcal{E}_r$ as a proportion of species in which reaction $r$ is essential among the species containing it in their metabolism. Then, $\mathcal{E}_r$ increases with reaction age $y_r$ as depicted in Fig.~\ref{fig:6}(a). This behavior suggests that older reactions are more likely to be important. After that, we consider the average local connectivity of reaction $r$ over different species containing it as a function of reaction age $y_r$, denoted as $\langle k_r\rangle(y_r)$. It was found that $\langle k_r\rangle$ tends to be large for old reactions [Fig.~\ref{fig:6}(b)]. Because old reactions are more likely to be essential as shown in Fig.~\ref{fig:6}(a), reactions connected to many other reactions are more likely to be essential than those connected to few, as confirmed weakly in Fig.~\ref{fig:6}(c). On the other hand, the lower mean connectivity of younger reactions allows us to understand the origin of low mean connectivity of old species [Fig.~\ref{fig:bm_ess}(c)]. 

Next, we introduce backup availablity (denoted as $\mathcal{B_A}(r)$) as the proportion of species that contain at least one backup partner of reaction $r$ among all species possessing reaction $r$. It is illustrated in Fig.~\ref{fig:6}(d) that the reactions with larger $\langle k_r\rangle$ are more likely to have higher backup availability. Given the positive correlation between average local connectivity $\langle k_r \rangle$ and reaction age $y_r$ in Fig.~\ref{fig:6}(a), one can then expect that old reactions will have higher backup availability than young reactions, as confirmed below.

%%%%%%%%%%%%%%%%%% new fig 7(old Fig. 11) fig:yrfEr %%%%%%%%%%%%
\begin{figure}
\centering
\includegraphics[width=0.99\columnwidth]{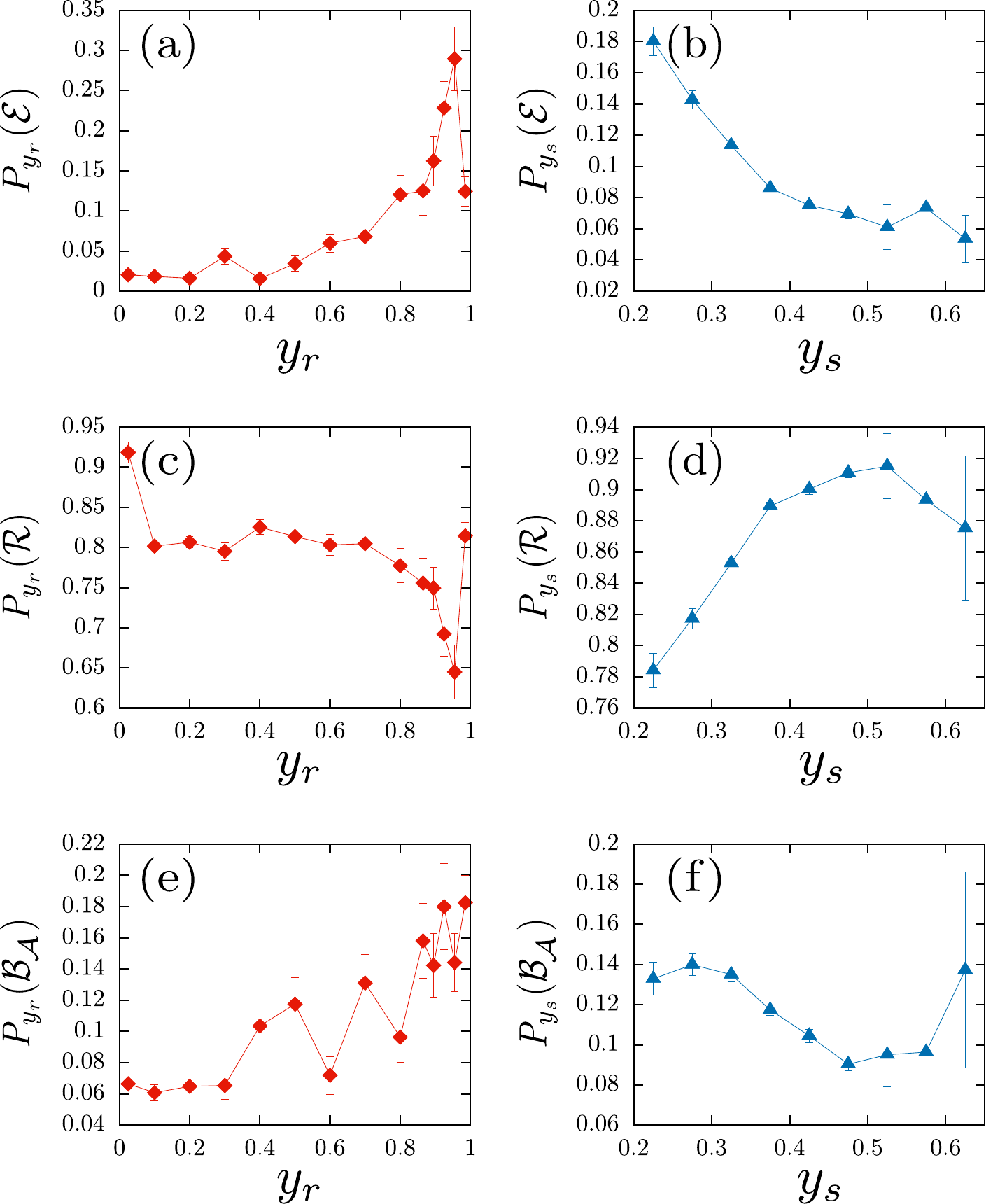}
\caption{The  probability of a reaction to be essential, redundant, and to have backup partners versus reaction age and species age. (a) Probability $P_{y_r}(\mathcal{E})$ of reaction $r$ in a species to be essential as a function of its age $y_r$. The PCC between $P_{y_r}(\mathcal{E})$ and $y_r$ is $0.244$. 
(b) Pobability $P_{y_s}(\mathcal{E})$ of a reaction in species $s$ to be essential as a function of species age $y_s$. They correlate negatively (PCC = $-0.718$).
(c) Probability $P_{y_r}(\mathcal{R})$ of reaction $r$ in a species to be redundant as a function of its age $y_r$.  They correlate negatively and the PCC is $-0.695$ with $P=0.005$.
(d) Probability $P_{y_s}(\mathcal{R})$ of a reaction in species $s$ to be essential as a function of species age $y_s$. They correlate positively (PCC = $0.661$). 
(e) Probability $P_{y_r}(\mathcal{B_A})$ that a reaction of age $y_r$ has one or more backup partners in a species. The PCC between $P_{y_r}(\mathcal{B_A})$ and $y_r$ is $0.119$. 
(f) Probability $P_{y_s}(\mathcal{B_A})$ that a reaction has one or more backup partners in a species of age $y_s$. The PCC between $P_{y_s}(\mathcal{B_A})$ and $y_s$ is negative as $-0.377$.}
\label{fig:yrfEr}
\end{figure}
%%%%%%%%%%%%%%%%%%%%%%%%%%%%%%%%%%%%%%%%%%%

From the observations that young species have a larger number of essential reactions [Fig.~\ref{fig:bm_ess}(b)] and that old reactions are more likely to be essential [Fig~\ref{fig:6}(a)], we infer that the likelihood of being essential will be large for old reactions located in young species' metabolic networks. 
Probabilities $P_{y_r} (\mathcal{E})$ and $P_{y_s} (\mathcal{E})$ that a reaction in a species is essential at reaction age $y_r$ and species age $y_s$, respectively, are shown in Fig.~\ref{fig:yrfEr}(a) and (b), and confirm this expectation. Note that $P_{y_r}(\mathcal{E})$ is related to $P_{y_r,y_s}(\mathcal{E})$ given in Fig.~\ref{fig:gradeDistribution} by the relation $P_{y_r}(\mathcal{E})= \sum_{y_s} P_{y_r, y_s}(\mathcal{E}) P_{y_r} (y_s)$ with $P_{y_r}(y_s)$ being the probability of a reaction of age $y_r$ to be present in a species of age $y_s$.

Such variations in essentiality depending on reaction age and species age reveal that there exist two major determinants of essentiality. One is the functional importance of a reaction in cellular metabolism: Old reactions are likely to be important, to contribute to the production of biomass components, and to be connected to many other reactions, resulting in the positive correlation between the probability of a reaction to be essential and reaction age. Most of nonessential reactions are redundant ones, as seen in Fig.~\ref{fig:gradePortion}. Therefore the probability of a reaction to be redundant appears to negatively correlate with the reaction age as shown in Fig.~\ref{fig:yrfEr}(c). The other determinant is related to the likelihood of development of alternative pathways, which varies from species to species. If inactivation of a reaction does not significantly affect the rest of the metabolism owing to the presence of alternative pathways, it will not remain essential but is classified here as active or redundant and may have backup partners. Alternative pathways are more likely to develop in old species that have had sufficient time for evolution. Therefore, the probability that a reaction is essential decreases with species age, whereas the probability to be redundant 
increases with species age, as demonstrated in Figs.~\ref{fig:yrfEr}(b) and \ref{fig:yrfEr}(d). 
Simultaneous inactivation of a pair of backup reactions may be considered for biomass generation; this notion leads us to the hypothesis that the likelihood that a reaction has a backup partner reaction manifests the same dependence on the reaction age and species age as essentiality does, as supported by  Figs.~\ref{fig:yrfEr}(e) and \ref{fig:yrfEr}(f).

\section{Summary and discussion}
 
Here we computationally investigated the lethality of inhibiting individual reactions in the metabolic networks of hundreds of species. The effect on biomass generation was used to quantify the lethality, which was found to depend on the organization of alternative pathways and the importance of the reaction. Our study revealed universal categorization: $\sim 10\%$ of the reactions in each species are essential, and the remaining 90\% are redundant, which may be an important characteristic of the risk management strategy of cellular metabolism against internal or external perturbations. We assigned evolutionary age to each reaction and species; this approach enables us to estimate when a reaction was introduced to the metabolism and to evaluate the degree of evolution of the metabolic network of each species. These parameters are useful for understanding the lethality landscape across reactions and species.

We found a significant negative correlation between the number of essential reactions in a species and its age; this result suggests that young reactions are abundant in old species and prevent local perturbations from affecting the metabolism globally, thereby helping to enhance the stability of cellular metabolism. 
At the level of individual reactions, the likelihood that a reaction in a species is essential was found to be large if the reaction is old, because old reactions are likely to be important and responsible for core functions in the metabolism, and small if the species is old, because the species is equipped with alternative pathways protecting important reactions. 
These results suggest that the metabolic network has evolved by recruiting new reactions and pathways to reduce the lethality of losing important reactions to some extent depending on the environmental requirements imposed on the species.
The cross-species characteristics of individual reactions obtained in our study can be considered intrinsic features and can have potential applications in medicine and drug discovery by allowing an investigator to classify and rank metabolic reactions in terms of their importance and stability.

This study can be extended in diverse directions for deeper understanding of the stability of metabolic networks. The use of an ideally rich medium for our FBA computation might have led us to a relatively small set of essential reactions in each species; larger sets of reactions could be identified as essential depending on the specific composition of various poor media. In addition to the FBA computation, the study of the pathways linking external compounds to the biomass components and the frequency with which a reaction participates in different pathways can offer another layer of knowledge on the essentiality of reactions. 

\begin{acknowledgments}
This work was supported by National Research Foundation of Korea (NRF) grants funded by the Korean government  [No. 2014-069005 (BK) and No. 2016R1A2B4013204 (DSL)], and the Center for Women in Science, Engineering and Technology (WISET) Grant funded by the Ministry of Science and ICT under the Program for Returners into R\&D (PK).
\end{acknowledgments}

%\section{Method}
\section*{Appendix A: Datasets and the exact expressions for the measured quantities}

\begin{figure}
\centering
\includegraphics[width=0.95\columnwidth]{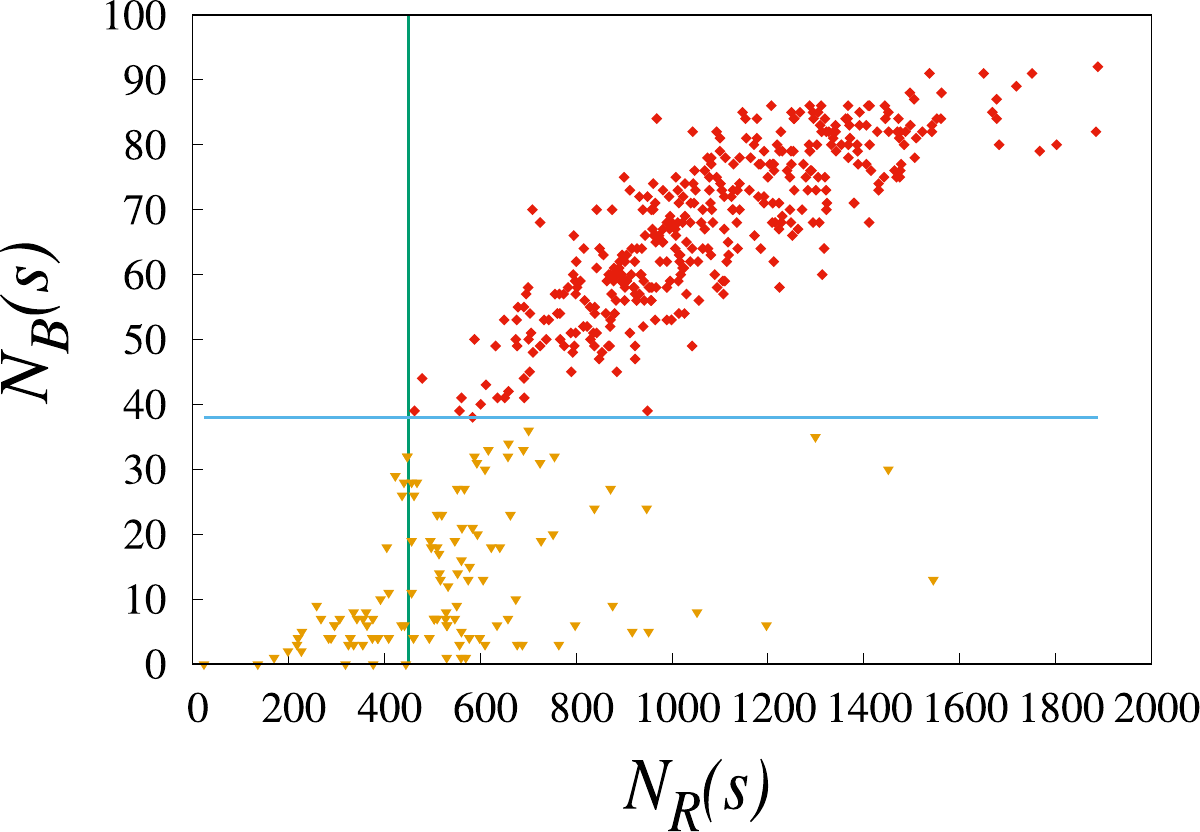}
\caption{Scattered data on species in the plane of  the number of biomass component $N_B(s)$ and the number of reactions $N_R(s)$ contained in each species. We find that 386 species in the region $N_B(s) \ge 38$ and $N_R(s) \ge 450$ are well separated from 120 species in the other region. Those 386 species have relatively large numbers of biomass components and reactions, and therefore features of metabolic networks of those species can be analyzed in this paper.      
}
\label{fig:OrgaSort}
\end{figure}

We constructed the metabolic networks in a bipartite form consisting of metabolic reactions $r$ and metabolites $c$ with adjacency matrix $A^{s,{\rm (bi)}}_{cr}=1$ or $0$ for 506 species by means of the BioCyc database, version 13.1. Let us denote the set of species we studied by $\mathbb{S}$. We considered set $\mathbb{B}$ of 129 biomass components identified in at least one of the following species {\it Bacillus subtilis}, {\it E. coli}, {\it Helicobacter pylori}, {\it Staphylococcus aureus}, {\it Methanosarcina barkeri}, and {\it Saccharomyces cerevisiae}. We checked whether each of them is generated in the metabolic network of each species by FBA to obtain the matrix $A^{(B)}_{sc}=1$ or $0$, indicating whether component $c$ is produced in the metabolic network of species $s$ in an ideally rich medium. The details of this FBA are described in the next section. As a result, we have the set $\mathbb{B}(s) = \{ b\in \mathbb{B}| A^{(B)}_{cs}=1\}$.  
Set $\mathbb{R}$ denotes the compilation of all 7919 reactions appearing in at least one species, and we have another matrix, the reaction$-$species matrix, $A_{rs}=1$ or $0$, indicating whether reaction $r$ appears in the metabolic network of species $s$. Therefore, we obtain the set of reactions $\mathbb{R}(s)=\{r\in \mathbb{R}|A_{rs}=1\}$ for each species $s$ and the set of species $\mathbb{S}(r) = \{s\in \mathbb{S}| A_{rs}=1\}$ for each reaction $r$. 
Given clear-cut stratification of species by the number of reactions $N_R(s)$ and the number of biomass components $N_B(s)$ as shown in Fig.~\ref{fig:OrgaSort}, we restrict our study to 386 species having $N_B(s) \ge 38$ and $N_R(s) \ge 450$.  In the text, we let $\mathbb{S}$ denote this set of 386 species. We find that $N_R=6911$ distinct reactions are found in at least one of these 386 species. 

%%%%%%%%%%%%%%%%%% Fig. 13 fig:OrgaSort %%%%%%%%%%%%
%%%%%%%%%%%%%%%%%%%%%%%%%%%%%%%%%%%%%%%%%%%

\section*{Appendix B: Details of FBA of 386 species and Biomasses}

To determine the essentiality grade of all nonisolated reactions, we perform FBA on the metabolic network of each species. 
The stoichiometric matrix for the computation is provided by the BioCyc database.
We assume that all the reactions are reversible, considering that most reactions can proceed in both directions under certain physiological conditions of temperature or metabolite concentrations~\cite{Helden,Croes2005,Nielsen}.
Assuming reversibility, we also add organism-specific exchange reactions that carry any metabolite outside the cytosol into the cytosol. 
This means that the organism may be able to absorb whatever it needs. 
Furthermore, we attach a reaction of organism-specific biomass synthesis to the metabolic network of an organism. 
For example, if the organism can synthesize metabolites A, B, and C among 129 biomass components, then the biomass reaction of the organism is  $A+B+C \xrightarrow{g_w} biomass$.  
An ideally rich medium is assumed to be available for every species' growth such that all the exchange reactions, including both uptake and secretion, can proceed without limit. This arrangement is actually realized in computation by setting the same upper bound for every exchange reaction. The flux of the biomass generation reaction is regarded as the growth rate of a given species $s$, and is denoted by $g_w(s)$. 
After that, we set the flux of each reaction $i$ to $0$ and perform FBA to obtain  $g_i(s)$, the growth rate with reaction $i$ inhibited.

\end{document}